# Gesture recognition with Brownian reservoir computing using geometrically confined skyrmion dynamics

## Authors

Grischa Beneke[1], Thomas Brian Winkler[1], Klaus Raab[1], Maarten A. Brems[1], Fabian Kammerbauer[1], Pascal Gerhards[2], Klaus Knobloch[2], Johan Mentink[3], Mathias Kläui[1,4]

[1] Institut für Physik, Johannes Gutenberg-Universität Mainz, 55128 Mainz, Germany
[2] Infineon Technologies Dresden, 01099 Dresden, Germany
[3] Radboud University, Institute for Molecules and Materials, 6525 AJ Nijmegen, the Netherlands
[4] Center for Quantum Spintronics, Norwegian University of Science and Technology, 7491 Trondheim, Norway

## Abstract

Physical reservoir computing (RC) is a beyond von-Neumann computing paradigm that harnesses the dynamical properties of a complex physical system (reservoir) to process information efficiently in tasks such as pattern recognition. This hardware-centered approach drastically reduces training efforts and holds potential for significantly reduced energy consumption operation. Magnetic skyrmions, topological, particle-like spin textures, are considered highly promising candidates for reservoir computing systems due to their non-linear interactions and established mechanisms for low power manipulation combined with thermally excited dynamics. So far spin-based reservoir computing has been used for static detection or has been based on intrinsic magnetization dynamics timescales, that require cumbersome rescaling of typically slower real-world data. Here we harness the power of time-multiplexed skyrmion RC by adjusting the intrinsic timescales of the reservoir to the timescales of real-world temporal patterns: we recognize hand gestures recorded with range-doppler radar on a millisecond timescale and feed the data as a time-dependent voltage excitation directly into our device. We observe the temporal evolution of the skyrmion trajectory using read-out at just one position of the reservoir, which allows for scaling onto the nanometer scale. We find that our hardware solution exhibits competitive performance compared with a state-of-the-art energy-intensive software-based neural network. The key advantage of our hardware approach lies in its capacity to seamlessly integrate data from sensors without the need for temporal conversion, thanks to the time-dependent input and tunable intrinsic timescales of the skyrmion dynamics in the reservoir. This feature enables real-time feeding of the reservoir, opening a multitude of applications.

## Introduction

Skyrmions are chiral magnetic whirls that have been shown to exhibit enhanced stability due to their non-trivial topology[1–4]. They have shown great potential in non-conventional computing devices[5–13], or as information carriers in novel data storage[3,14–16]. Stabilized due to the Dzyaloshinskii-Moriya interaction, they can be present in bulk systems[17], or in thin film systems[2,18–20], where they exhibit particle-like behavior[3,5]. Proposals for applications exploit the unique characteristics of skyrmions for instance as bit-like information carriers[14,15], and use deterministic motion, controlled nucleation and annihilation of the spin textures for deterministic memory operations. On the other hand, stochastic (thermal) dynamics is exploited for Brownian computing approaches[8,13,21,22]. In Brownian computing, the computation speed is tied to the systems' diffusion coefficient and can thus be tuned[8,23]. Therefore, skyrmion systems are particularly favorable as skyrmions can exhibit significant thermal diffusion at room temperature[5] and diffusion-tuning mechanisms have been developed recently that allow for tuning the thermal dynamics over many orders of magnitude[23]. In both deterministic and stochastic applications, the manipulation of skyrmions can be achieved efficiently even with ultra-low current densities through currents that generate spin transfer torques[24,25] and spin-orbit torques[26,27], as well as external fields[28], strain[29] and temperature gradients[30]. In thermally-activated skyrmion systems, pinning effects can be overcome by thermal excitations allowing for directed motion at even lower current densities[5,21,31] where ultra-low currents bias the diffusive dynamics[21]. Thereby, the Brownian computing paradigm can offer low-power computing while simultaneously overcoming reproducibility issues due to variations between devices.

Physical Reservoir Computing (RC) represents a key machine learning approach, where a non-linear physical reservoir is harnessed to process and project input data into a higher-dimensional state-space where they become for instance linearly seperable[32]. By performing a measurement of the reservoir, the complex state is subsequently mapped to an output state of measurement values. Using this concept, one can potentially reduce a highly complex problem to a linear one, given the dynamics of the system are suited to process the signal properly. This has a major advantage compared to deep recurrent neuronal networks where all the weights are trained, which is complex, slow and energy inefficient. In RC only the output weights are trained while the reservoir is fixed, resulting in faster and lower energy operation[33]. Further, as the computation is performed by the systems' efficient intrinsic dynamics, the power consumption of such a device is typically much lower than inferring a software-based solution with similar performance.

Reservoir computing has been studied in magnetic systems[9–13,21,34–36] for static spatially multiplexed pattern recognition[21] and for the recognition of dynamic time-varying signals using spin structure dynamics[9], spin torque oscillators[37] and spin waves[34], which have intrinsic timescales in the MHz to GHz regime. While a proof-of-concept has been demonstrated for vowel recognition[38], the involved intrinsic magnetic dynamic frequencies in the MHz-GHz range require a complex and energy intensive rescaling of the timescales of the spoken language or many other real-world signals that are to be recognized.

So most conspicuous applications for dynamic pattern recognition range from speech to the detection of motion and such dynamics is occurring on the μs-s timescale, which is however orders of magnitude slower than the spin dynamics conventionally used for reservoir computing requiring cumbersome and energy intensive rescaling of the timescales.

Here we employ thermally activated diffusion and current-induced displacement of skyrmions in a reservoir to detect real human gestures. We input the doppler radar data of gestures as ultra-low power current-driven dynamics that in combination with geometrical confinement results in an intrinsic zero-power reset mechanism. Having comparable time scales of the doppler radar data and the intrinsic dynamics and thus processing speed of the reservoir enables direct feeding of the sensor data into the reservoir. We find that the radar data of different hand gestures is detected in our hardware reservoir with a fidelity that is at least as good as a state-of-the-art software-based neural network approach.

## Results

Our device consists of a $Ta(5)/Co_{20}Fe_{60}B_{20}(0.95)/Ta(0.09)/MgO(2)/Ta(5)$ multilayer stack (thickness of layers given in nanometers in parentheses) that is structured into an equilateral triangular geometry, with a side length of 36 μm. Figure 1 schematically depicts such a device. Such geometrical confinement was shown to lead to an equilibrium position of the skyrmion in the center[16,39–41] combined with non-linear current-driven skyrmion displacements due to spin-orbit torques[21,26,27,42–45].

The thermally activated diffusive dynamics results in stochastic motion, due to the low pinning of the multilayer stack[5,46,47] allowing for displacement at very small current densities and leading to a zero-power reset mechanism when the current is switched off. To tune the speed of the dynamics, an additional oscillating out-of-plane magnetic field component is used, that was shown to tune the diffusion speed over orders of magnitude[23].

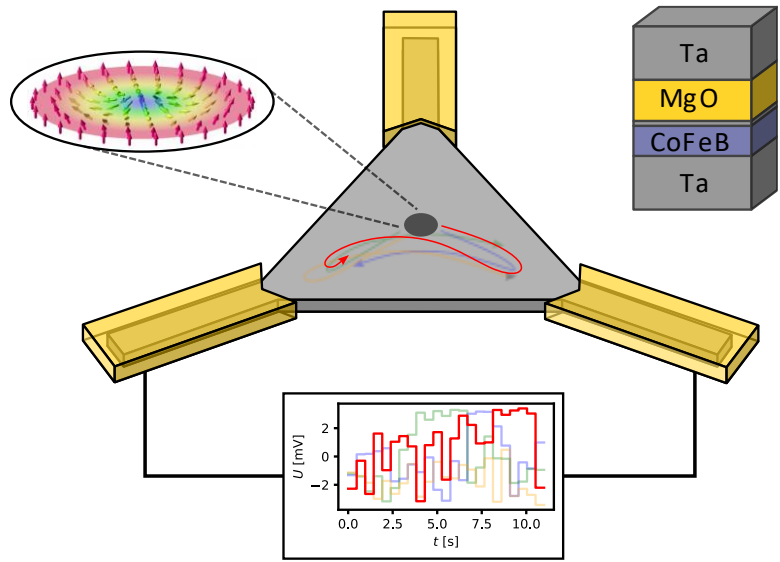

*Figure 1: Schematic representation of the dynamic Brownian reservoir computer. The triangular stack (light grey; composition shown in the top right) is connected with gold/chromium contacts at the corners (half transparent yellow). In the used setup, only the lower corners are connected to the time-dependent voltage. The skyrmion is presented as a dark gray spot as visualized by magneto-optical Kerr microscopy. Possible skyrmion response trajectories to the time dependent voltages are shown by colored arrows. Inset adapted from Karin Everschor-Sitte and Matthias Sitte, licensed under Creative Commons Attribution-Share Alike 3.0 Unported (CC BY-SA 3.0).*

The skyrmion is nucleated by using an in-plane magnetic field pulse, while an additional constant out-of-plane magnetic field stabilizes the skyrmion[5,21,23]. After the skyrmion is nucleated, the edge repulsion along with thermal dynamics cause the skyrmion to reside in the center of the device. Consequently, whether upon initialization or subsequent excitation through current-induced motion, the device inherently reverts to its original ground state due to this automatic reset mechanism[41].

To operate as a time-multiplexed reservoir, different signals in the form of time-dependent voltages are applied to the device. This results in a time-dependent displacement of the skyrmion. As the magnitude of the applied voltage increases, and subsequently the current density and thus SOT, the skyrmion position is pushed closer towards the corners of the device. This relationship is non-linear due to the non-linearity of the boundary interaction potential in our system[48]. The skyrmion is driven by very low current densities up to $2.9 \times 10^7 \ Am^{-2}$ at half width of the triangle. This current density is lower than any previously demonstrated displacement in multilayer skyrmion stacks is realized by the artificial mitigation of pinning effects by the oscillating magnetic out-of-plane field[23]. Stronger current densities may result in the annihilation of the skyrmion, while low current densities cause motion indistinguishable compared to purely thermal movement. The amplitude of the input signal is scaled within the range of negative and positive maximum voltage, enabling displacement of the skyrmion to either the left or right corner of the device (see methods section for more details). After finishing the input of the radar signal data, the skyrmion relaxes back to the center.

With this procedure, every possible time dependent voltage can be mapped to a set of time dependent movements of the skyrmion. Due to the super-imposed motion caused by current induced SOT and thermally activated diffusion, which enables skyrmions to overcome pinning sites, the skyrmion response will not be identical, but similar enough for the reservoir to function reliably. An essential requirement of reservoir computing is the non-linear reservoir. The presented skyrmion reservoir is non-linear in multiple aspects, the skyrmion velocity being dependent on the current[44], the current density being dependent on the location, the skyrmion edge repulsion[48], and the skyrmion diffusion.

In our proof-of-concept experiment, the time-dependent position of the skyrmion is captured using a magneto-optical Kerr-effect microscope that captures the full information about the position of the skyrmion at a framerate of $16\ Hz$. In real applications, a single time dependent readout of a magnetic tunnel junction (MTJ) would be used, which is emulated here by reading the data from a fixed $0.5\ \mu m$ diameter area that is positioned in the device. A detailed explanation of the conversion process from Kerr images to MTJ signals is provided in the methods section and is depicted in Fig. 2a and b. Moreover, depending on the input timescale, we reduce the time resolution by using only every $n^{th}$ measurement frame, for which n is out of [16, 8, 4, 2, 1].

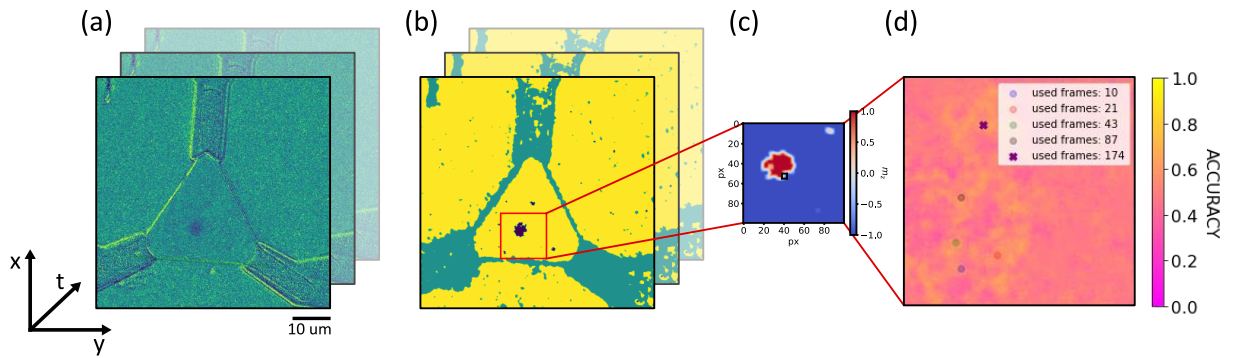

*Figure 2: Data processing.* a) Kerr-microscope video of the device a with skyrmion inside. b) U-Net prediction to reduce noise. Skyrmion prediction in purple, contacts and defects in turquoise, background and magnetic material in yellow. The skyrmion prediction is used to simulate the artificial MTJs. c) Analyzed video part with one possible emulated MTJ marked. d) Validation accuracy map of SVM for different input pixels. In this case for differentiating gesture all gestures with 87 data points in time. Locations with the highest accuracy for different number of data points in time are marked. As the emulated MTJs have a diameter of $0.5\ \mu m$, only the center is marked.

To demonstrate the device's capability, the input comprises gesture data, which is processed to distinguish various human hand gestures, encoded as class numbers (see Fig 3a). The doppler radar-captured data is fed into the skyrmion reservoir, as elaborated in the methods section, such that each distinct hand gesture corresponds to a unique 23 frames long time-dependent voltage signal. An example can be seen in Fig. 1. The experimentally used subset of the dataset of 4800 recorded gestures employed in this study comprises four distinct hand gestures, with each gesture being repeated 50 times, thereby yielding a cumulative total of 200 recorded instances. While the large skyrmions and their dynamics used here are motivated by the limited resolution of the optical read-out, the use of MTJs for the read-out will allow for scaling of both the spatial device size as well as the dynamic timescale over a broad range to detect many types of signals.

To find the best position of the readout MTJ on the device, every possible position of the MTJ is evaluated independently. The emulated MTJs always correspond to multiple pixels that match their size. We train a linear support vector machine (SVM) on each position to see which the best is to differentiate the different gestures. Note that this optimization must only be performed once. The optimization procedure produces an accuracy map of the device, see Fig. 2, which also illustrates all previous steps. Each time a state-of-the-art linear SVM is trained while K-Folding with $K = 10$ is used for robust evaluation of the validation set performance.

Figure 3 shows the benchmark comparison for our reservoir compared to the state-of-the-art software detection. Comparing the detection accuracy of different gestures, we find a very competitive performance of the reservoir. Especially for gesture pairs 01, 12 and 13 we find even an improvement of the skyrmion reservoir compared to a linear SVM and other methods. Here the best performing number of data points were chosen for the reservoir accuracy (see methods section for more information). It is also visible that the non-linear transformation to improve the skyrmion response does not change the accuracy significantly for the pure software-based approach.

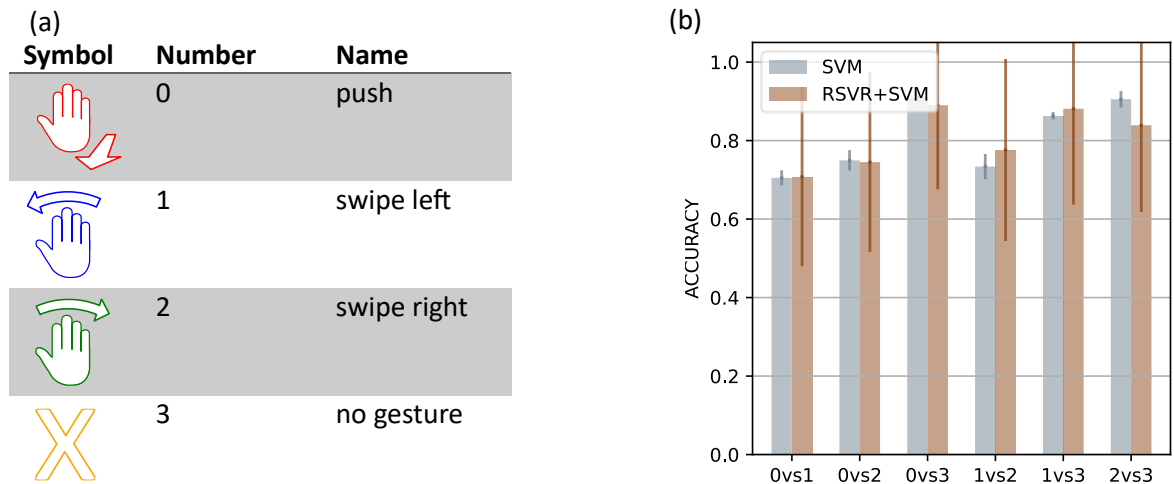

*Figure 3:* a) Table of different hand gestures in data set. b) Comparison detection accuracy between SVM and reservoir plus SVM for different gesture pairs. The error is calculated with the standard deviation of different SVM resulting from the K-Folding.

## Discussion

We successfully demonstrated that our skyrmion reservoir can be used as a time-multiplexed reservoir computer for time series classification problems, exemplified on range-doppler gesture data. In the pairwise classification, the Reservoir is competitive and can even exhibits superior performance compared to state-of-the-art software-based classification methods. We note that while the performance is already competitive with a read-out of the reservoir based on a single MTJ, one can easily improve our results by using multiple MTJs as readouts, thus capturing more regions of the skyrmion trajectories. However, we stress here the fact that even the simplest case of a single MTJ readout is already providing excellent performance.

We note that a key advantage of our approach is that the device can be adapted to other tasks easily. The complexity and recognition accuracy can be optimized using a different device layout or employing multiple skyrmions. Furthermore, the timescale of the thermal dynamics that governs the timescale of the detected signals can be tuned by a small AC excitation over multiple orders of magnitude[23]. And scaling the skyrmion reservoir down to the nanoscale, is a means to enhance energy efficiency and operational speed, as discussed in the Supplementary Material. The rationale behind this lies in the reduction of the skyrmions' displacement time and current-induced motion's dependence on current density, smaller devices, both favoring smaller devices, which thus, need less energy while additionally allowing for a larger device density. Coupled with the reduced displacement time of skyrmions within smaller devices, this scaling could be used as a variable timescale and ultra-low energy reservoir computing implementation, offering promising prospects for future applications.

## Methods

**Sample Preparation**

The multilayer stack employed in this study consists of the following layers Ta(5)/Co$_{20}$Fe$_{60}$B$_{20}$(0.95)/Ta(0.09)/MgO(2)/Ta(5), where the values enclosed in parentheses indicate the respective thickness in nanometers, and the subscripted numbers represent the concentration in percentage. The stack is deposited using a Singulus Rotaris magnetron sputtering system. The sample exhibits perpendicular magnetic anisotropy (PMA) due to the interface between the ferromagnetic Co$_{20}$Fe$_{60}$B$_{20}$ and the MgO. The Ta(0.09) interlayer adjusts the PMA and generates a low-energy landscape for magnetic skyrmions. The $5\ nm$ thick capping layer consisting of tantalum prevents oxidation, together resulting in a multilayer stack that hosts skyrmions at above room temperature experiencing low pinning and thermal diffusion.

The Raith Electron Beam Pioneer system was employed for electron beam lithography in order to fabricate patterned structures. The sample underwent etching with Argon ions through the IonSys Model 500 ion beam etching system.

The device consists of an equilateral triangle with an edge length of $36\ \mu m$. At the corners extra rectangles are attached with a width of $4\ \mu m$ and length of $15\ \mu m$ for better contact connectivity. These contacts are composed of a $5\ nm$ layer of chromium overlaid with $60\ nm$ gold and were fabricated utilizing electron beam lithography in conjunction with a lift-off technique. The device under examination exhibits a measured resistance of $1.1\ kOhm$ between two contacts.

**Measurement Setup**

To visualize the skyrmions, the experimental setup employs a microscope manufactured by evico magnetics GmbH, which leverages the magneto-optical Kerr effect. The microscope is connected to a CCD camera recording at 16 frames per second with an exposure time of $62.5\ ms$ and a resolution of $1344 \times 1024$ pixel. With a 2x2 binning, the resolution is reduced to $672 \times 512$ pixels in favor of a higher signal-to-noise ratio. The camera captures an area of $80 \times 61\ \mu m$. To achieve better contrast a differential image between the skyrmion state and the saturated state is used.

The setup consists of custom-designed coils, featuring one coil for generating the out-of-plane (OOP) magnetic field and an additional pair for the in-plane (IP) field. To the OOP coil, a Peltier element is attached, giving an achievable temperature range from $285 - 360\,K$, with a thermal stability of $0.3\,K$. In the used setup a temperature of $325\,K$ was chosen, controlled with an PT100 resistive heat sensor. Furthermore, to further enhance thermal stability, the entire microscope system is housed within a laminar flow enclosure equipped with precise temperature control. This strategic implementation results in minimal thermal drift within the experimental setup, facilitating extended measurement durations at higher magnification levels.

The skyrmion is nucleated with an IP field pulse of $20\,mT$ and a stabilizing OOP bias field of $100\,\mu T$. After nucleation the additional OOP fields oscillations are started with an amplitude of $50\,\mu T$ and a frequency of $100\,Hz$, while keeping a bias of $100\,\mu T$. As the input signal, a time dependent voltage is applied to two corners using an arbitrary waveform generator. The recorded imagery is synchronized to the waveform generator.

**Data Evaluation**

The dataset consists of 4800 measurements of 4 classes of gestures: "0: push", "1: swipe left", "2: swipe right" and "3: no gesture" and was already successfully used to train spiking neural networks[49]. Varying people stood at the same position in regard to a radar sensor and performed hand gestures according to the classes. Every gesture measurement consists of 23 frames taken in real-time. The motion was detected by two Infineon Technologies radars sensors of the type BGT60TR13C. The data has to be transformed by exploiting micro Doppler effects[50], and obtaining Doppler maps, as described in Ref. [51] and Ref. [52]. Mainly, Fourier-transformations are performed to obtain the Doppler frequency shifts, resulting in two-channel maps (Range-Doppler and Range-Angle), showing the angle or the amplitude against the relative velocity and the distance to the radars. Figure SUP1 shows one exemplary "0 − push" gesture after the Range-Dopler (RD) and the Range-Angle (RA) transformation.

As we want to minimize the data fed into the reservoir, we restrict ourselves to only one voxel of the maps. To find the one containing the most information, we perform multi-class radial basis functions Support Vector Machine (rbfSVM)[53,54] classification on every possible index (Range-Doppler/Range-Angle), to find the input voxel (23 consecutive time-steps of one particular index) that performs best on all 4 gestures at once. We further checked if data pooling (Average of 1x1, 2x2, or 4x4 tiling) increases classification. Every input was trained with K-Folding and $K = 10$ to obtain statistically robust values. The best score was obtained by averaging the performance of the $K$ test sets. The respective voxel is marked with a red box in Figure SUP1. Reducing the data to one input is also required to obtain experimental data in a reasonable time, as the manual experiments are time-consuming and must be taken with care. In a non-proof-of-concept device, this training process could of course be automated using specifically tailored hardware. The pairwise results shown in the main text are obtained by the same voxel, namely the one that performed best.

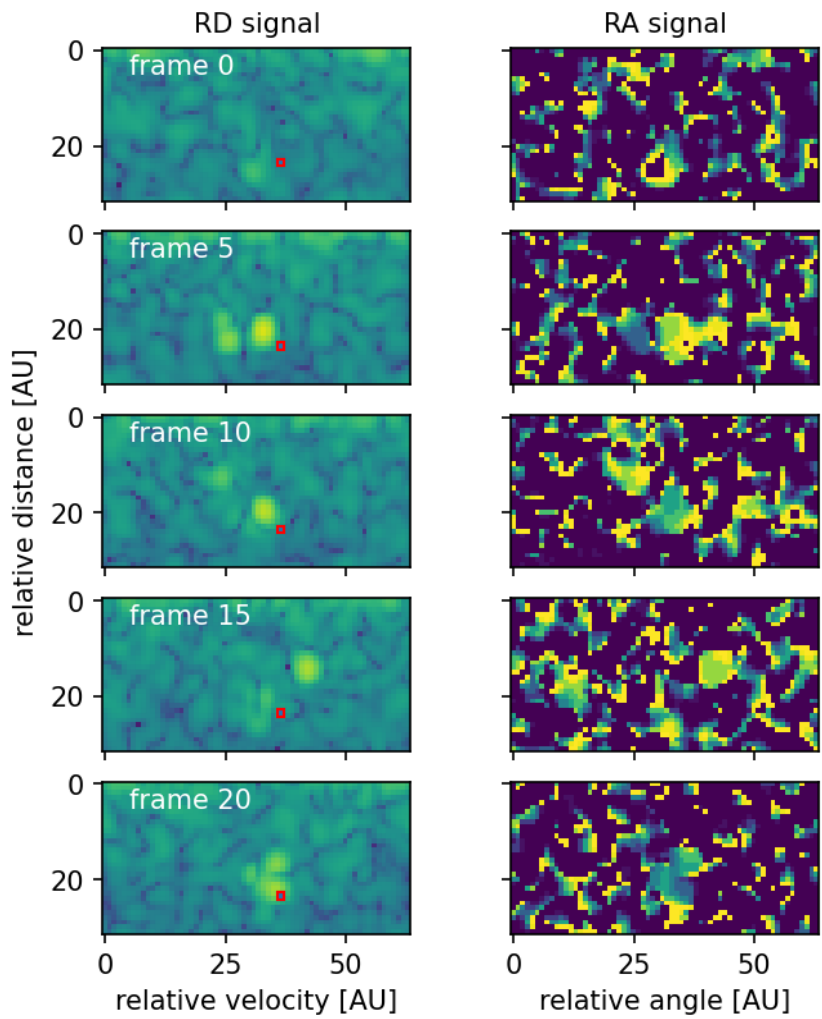

**Figure SUP1:** Range-Doppler and Range-Angle maps for one specific gesture and one-time frame. Apart from noise, one can detect two blobs in the maps, one referring to the person's body gesticulating, and the other to the hand doing the movement. We also marked the voxel with a red box that was used as input for our Skyrmion Reservoir.

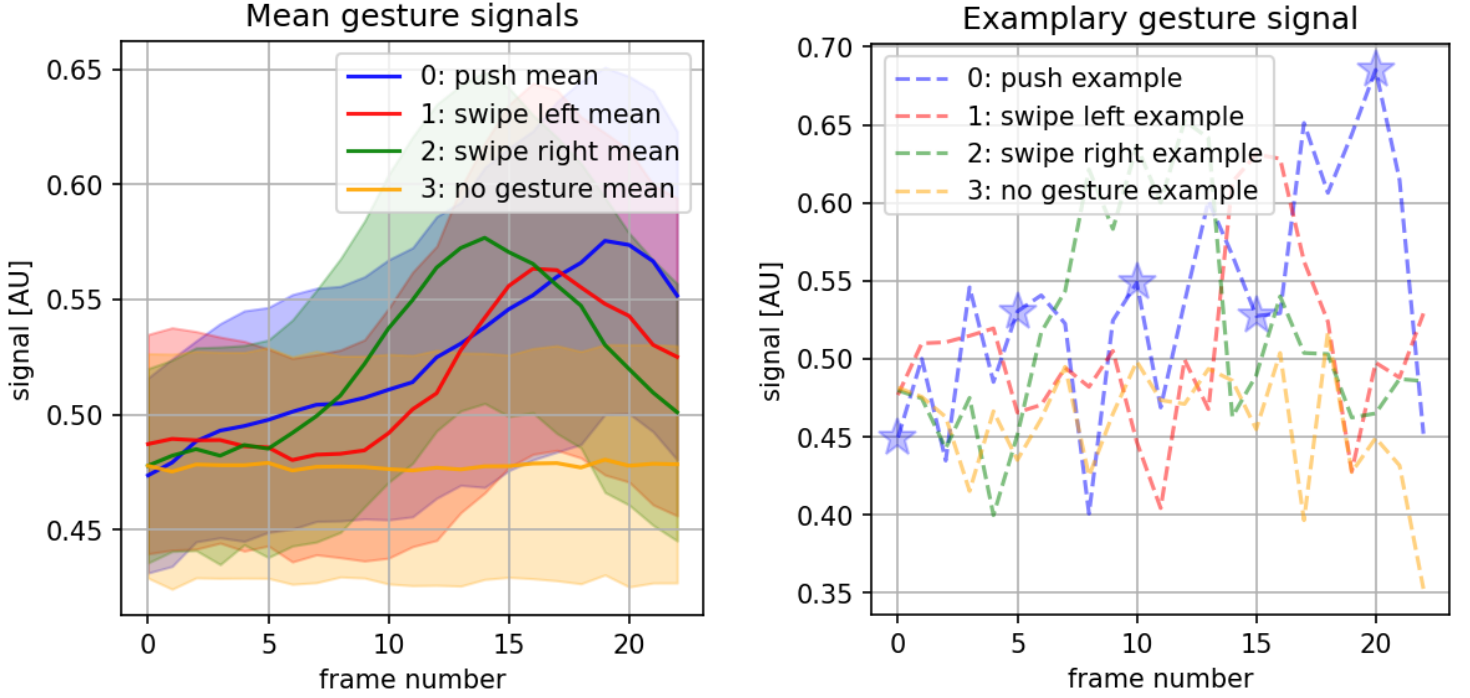

**Figure SUP2** *Left: We show the normalized mean signal of every gesture, for the voxel performing best in the Range-Doppler transformed data. Right: Exemplary gestures of every class. The blue star marks the values which are shown in red boxes in Fig. SUP1, which shows the same gesture.*

After the input data was defined, signal-and-time matching was necessary to transform the arbitrary input signal into a current density that significantly moves the skyrmion without annihilating it. In addition, the time scales of the input need to match, e.g. the input signal should vary at similar timescales as the skyrmion dynamics happens. The matching functions read as:

$$U(t) = c_1 \cdot \tanh((v(t) - MEAN) \cdot c_2)$$

With $v(t)$ being the signal at a specific time frame $t$, $MEAN$ the mean value of the dataset (to normalize them around zero. The other parameters were chosen with $c_1 = 3.5\ mV$ and $c_2 = 4$ in such a way that the skyrmions are driven far enough out of equilibrium to distinguish trajectories, but not too much that there is no significant probability for a skyrmion to be annihilated at the boundary. The 23 timeframes were mapped to an input. After each measurement, we let the skyrmion relax back to the center position, which depending on the diffusion coefficient takes milliseconds to seconds. Also, the gesture order was randomized to reduce the possibility of any bias regarding potential insufficient relaxation.

As typically some threshold current strength needs to be overcome to move the skyrmion at all, we also use the tanh function to effectively increase the normalized input.

After the MOKE measurements were performed, the data was cropped to 174 timeframes per gesture. The contrast was then enhanced using the 2.5% outer percentiles for adjustment, after all, artifacts (that might occur due to the combination of background-subtraction and sample-drift) were set to the mean value of each gesture. Then, to detect the skyrmion, a Convolutional Neural network was used which can reliably segment the image into the skyrmion (label: 0), defect (1), and magnetic background (label: 2) [55], while for our analysis we reduce the labels to (Skyrmion:1 and Background/defects:0).

To mimic the MTJ-readout, we chose a rectangular size of 5x5 pixels, while the MOKE resolution is $\approx$ 120 $nm/pixel$. An MTJ of this size should easily be manufactured[56]. We average the segmented label in that area to obtain more than two discretizations of the output space and interpret the value as a normalized resistance of the MTJ (0-lowest resistance – 1 highest resistance or vice versa). We use every possible position (MTJ is shifted by 1 pixel in each direction) in the confinement as input for an SVM that is part of the reservoir computing and obtain accuracy maps for the reservoir. We also check how many time frames are optimal, while we scan a range of every n$^{th}$ timeframe, with $n \in [16, 8, 4, 2, 1]$ and again K-Folding with $K = 10$, while the benchmarking was chosen to be the maximum value of the average accuracies of the test sets. We compare the outputs of these SVMs in the main text, Fig 3.

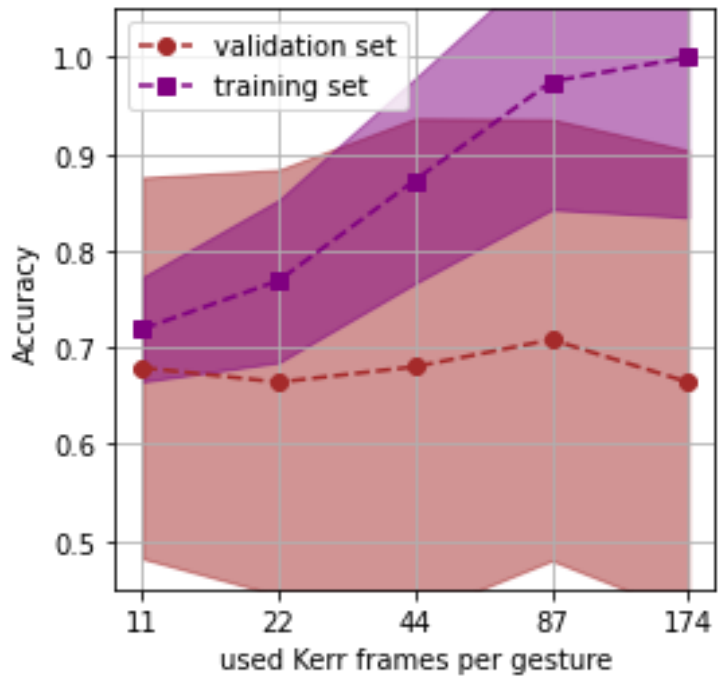

*Figure SUP 3* *Performance of the Reservoir, when different amount of Kerr video frames is used for the evaluation. The plots show the pairwise gesture comparison 0-push vs 1-swipe left. We see that the performance on the validation set peaks at specific amount of input (in this case 0.708 for 87 data points), while above we enter an overfitting regime of the linear SVM. For the plots in Fig 3 in the main text, we plot the best performance we find in the validation set. The best MTJ position for varying input size is also indicated in Fig 2 d) in the main text.*

## Acknowledgements
The gesture data was kindly provided by Infineon for use in this research.

The work in Mainz (G.B., T.B.W., K.R., M.A.B., F.K., and M.K.) was supported by the Deutsche Forschungsgemeinschaft (DFG, German Research Foundation) projects 403502522 (SPP 2137 Skyrmionics), 49741853, and 268565370 (SFB TRR173 projects A01 and B02) as well as TopDyn and the Zeiss foundation through the Center for Emergent Algorithmic Intelligence. J.H.M. acknowledges funding the Shell-NWO/FOM-initiative "Computational sciences for energy research" of Shell and Chemical Sciences, Earth and Life Sciences, Physical Sciences, FOM, and STW. The work is a highly interactive collaboration supported by the Horizon 2020 Framework Program of the European Commission under FETOpen grant agreement no. 863155 (s-Nebula) and ERC-2019-SyG no. 856538 (3D MAGiC) and the Horizon Europe project no. 101070290 (NIMFEIA), which M.K. and J.H.M. acknowledge. M.A.B. thanks the DFG TRR146 for partial financial support. We would additionally like to acknowledge helpful discussions with Peter Virnau.

# Bibliography

1. Bogdanov, A. & Hubert, A. Thermodynamically stable magnetic vortex states in magnetic crystals. *J. Magn. Magn. Mater.* **138**, 255–269 (1994).

2. Jiang, W. *et al.* Skyrmions in magnetic multilayers. *Phys. Rep.* **704**, 1–49 (2017).

3. Everschor-Sitte, K., Masell, J., Reeve, R. M. & Kläui, M. Perspective: Magnetic skyrmions—Overview of recent progress in an active research field. *J. Appl. Phys.* **124**, 240901 (2018).

4. Je, S.-G. *et al.* Direct Demonstration of Topological Stability of Magnetic Skyrmions via Topology Manipulation. *ACS Nano* **14**, 3251–3258 (2020).

5. Zázvorka, J. *et al.* Thermal skyrmion diffusion used in a reshuffler device. *Nat. Nanotechnol.* **14**, 658–661 (2019).

6. Zhang, X., Ezawa, M. & Zhou, Y. Magnetic skyrmion logic gates: conversion, duplication and merging of skyrmions. *Sci. Rep.* **5**, 9400 (2015).

7. Pinna, D. *et al.* Skyrmion Gas Manipulation for Probabilistic Computing. *Phys. Rev. Appl.* **9**, 064018 (2018).

8. Brems, M. A., Kläui, M. & Virnau, P. Circuits and excitations to enable Brownian token-based computing with skyrmions. *Appl. Phys. Lett.* **119**, 132405 (2021).

9. Msiska, R., Love, J., Mulkers, J., Leliaert, J. & Everschor-Sitte, K. Audio Classification with Skyrmion Reservoirs. *Adv. Intell. Syst.* **5**, 2200388 (2023).

10. Pinna, D., Bourianoff, G. & Everschor-Sitte, K. Reservoir Computing with Random Skyrmion Textures. *Phys. Rev. Appl.* **14**, 054020 (2020).

11. Prychynenko, D. *et al.* Magnetic Skyrmion as a Nonlinear Resistive Element: A Potential Building Block for Reservoir Computing. *Phys. Rev. Appl.* **9**, 014034 (2018).

12. Bourianoff, G., Pinna, D., Sitte, M. & Everschor-Sitte, K. Potential implementation of reservoir computing models based on magnetic skyrmions. *AIP Adv.* **8**, 055602 (2018).

13. Lee, O. *et al.* Perspective on unconventional computing using magnetic skyrmions. *Appl. Phys. Lett.* **122**, 260501 (2023).

14. Fert, A., Cros, V. & Sampaio, J. Skyrmions on the track. *Nat. Nanotechnol.* **8**, 152–156 (2013).

15. Parkin, S. & Yang, S.-H. Memory on the racetrack. *Nat. Nanotechnol.* **10**, 195–198 (2015).

16. Zhang, X. *et al.* Skyrmion-skyrmion and skyrmion-edge repulsions in skyrmion-based racetrack memory. *Sci. Rep.* **5**, 7643 (2015).

17. Tokura, Y. & Kanazawa, N. Magnetic Skyrmion Materials. *Chem. Rev.* **121**, 2857–2897 (2021).

18. Dohi, T., Reeve, Robert. M. & Kläui, M. Thin Film Skyrmionics. *Annu. Rev. Condens. Matter Phys.* **13**, annurev-conmatphys-031620-110344 (2022).

19. Finocchio, G., Büttner, F., Tomasello, R., Carpentieri, M. & Kläui, M. Magnetic skyrmions: from fundamental to applications. *J. Phys. Appl. Phys.* **49**, 423001 (2016).

20. Wiesendanger, R. Nanoscale magnetic skyrmions in metallic films and multilayers: a new twist for spintronics. *Nat. Rev. Mater.* **1**, 16044 (2016).

21. Raab, K. *et al.* Brownian reservoir computing realized using geometrically confined skyrmion dynamics. *Nat. Commun.* **13**, 6982 (2022).

22. Jibiki, Y. *et al.* Skyrmion Brownian circuit implemented in continuous ferromagnetic thin film. *Appl. Phys. Lett.* **117**, 082402 (2020).

23. Gruber, R. *et al.* 300-Times-Increased Diffusive Skyrmion Dynamics and Effective Pinning Reduction by Periodic Field Excitation. *Adv. Mater.* **35**, 2208922 (2023).

24. Jonietz, F. *et al.* Spin Transfer Torques in MnSi at Ultralow Current Densities. *Science* **330**, 1648–1651 (2010).

25. Yu, X. Z. *et al.* Skyrmion flow near room temperature in an ultralow current density. *Nat. Commun.* **3**, 988 (2012).

26. Hrabec, A. *et al.* Current-induced skyrmion generation and dynamics in symmetric bilayers. *Nat. Commun.* **8**, 15765 (2017).

27. Litzius, K. *et al.* Skyrmion Hall effect revealed by direct time-resolved X-ray microscopy. *Nat. Phys.* **13**, 170–175 (2017).

28. Chen, Y. *et al.* Nonlinear gyrotropic motion of skyrmion in a magnetic nanodisk. *J. Magn. Magn. Mater.* **458**, 123–128 (2018).

29. Sun, Y. *et al.* Experimental demonstration of a skyrmion-enhanced strain-mediated physical reservoir computing system. *Nat. Commun.* **14**, 3434 (2023).

30. Yu, X. *et al.* Real-space observations of 60-nm skyrmion dynamics in an insulating magnet under low heat flow. *Nat. Commun.* **12**, 5079 (2021).

31. Gruber, R. *et al.* Skyrmion pinning energetics in thin film systems. *Nat. Commun.* **13**, 3144 (2022).

32. Tanaka, G. *et al.* Recent advances in physical reservoir computing: A review. *Neural Netw.* **115**, 100–123 (2019).

33. Lukoševičius, M. & Jaeger, H. Reservoir computing approaches to recurrent neural network training. *Comput. Sci. Rev.* **3**, 127–149 (2009).

34. Lee, M.-K. & Mochizuki, M. Handwritten digit recognition by spin waves in a Skyrmion reservoir. *Sci. Rep.* **13**, 19423 (2023).

35. Misba, W. A. *et al.* Spintronic Physical Reservoir for Autonomous Prediction and Long-Term Household Energy Load Forecasting. *IEEE Access* **11**, 124725–124737 (2023).

36. Yokouchi, T. *et al.* Pattern recognition with neuromorphic computing using magnetic field–induced dynamics of skyrmions. *Sci. Adv.* **8**, eabq5652 (2022).

37. Torrejon, J. *et al.* Neuromorphic computing with nanoscale spintronic oscillators. *Nature* **547**, 428–431 (2017).

38. Romera, M. *et al.* Vowel recognition with four coupled spin-torque nano-oscillators. *Nature* **563**, 230–234 (2018).

39. Iwasaki, J., Mochizuki, M. & Nagaosa, N. Current-induced skyrmion dynamics in constricted geometries. *Nat. Nanotechnol.* **8**, 742–747 (2013).

40. Pepper, R. A. *et al.* Skyrmion states in thin confined polygonal nanostructures. *J. Appl. Phys.* **123**, 093903 (2018).

41. Song, C. *et al.* Commensurability between Element Symmetry and the Number of Skyrmions Governing Skyrmion Diffusion in Confined Geometries. *Adv. Funct. Mater.* **31**, 2010739 (2021).

42. Jiang, W. *et al.* Blowing magnetic skyrmion bubbles. *Science* **349**, 283–286 (2015).

43. Litzius, K. *et al.* The role of temperature and drive current in skyrmion dynamics. *Nat. Electron.* **3**, 30–36 (2020).

44. Woo, S. *et al.* Observation of room-temperature magnetic skyrmions and their current-driven dynamics in ultrathin metallic ferromagnets. *Nat. Mater.* **15**, 501–506 (2016).

45. Woo, S. *et al.* Spin-orbit torque-driven skyrmion dynamics revealed by time-resolved X-ray microscopy. *Nat. Commun.* **8**, 15573 (2017).

47. Kerber, N. *et al.* Anisotropic Skyrmion Diffusion Controlled by Magnetic-Field-Induced Symmetry Breaking. *Phys. Rev. Appl.* **15**, 044029 (2021).

48. Ge, Y. *et al.* Constructing coarse-grained skyrmion potentials from experimental data with Iterative Boltzmann Inversion. *Commun. Phys.* **6**, 30 (2023).

49. Kreutz, F., Gerhards, P., Vogginger, B., Knobloch, K. & Mayr, C. G. Applied Spiking Neural Networks for Radar-based Gesture Recognition. in *2021 7th International*

*Conference on Event-Based Control, Communication, and Signal Processing (EBCCSP)* 1–4 (2021). doi:10.1109/EBCCSP53293.2021.9502357.

50. Dekker, B. *et al.* Gesture recognition with a low power FMCW radar and a deep convolutional neural network. in *2017 European Radar Conference (EURAD)* 163–166 (IEEE, 2017). doi:10.23919/EURAD.2017.8249172.

51. Gamba, J. *Radar Signal Processing for Autonomous Driving*. (Springer Singapore, 2020). doi:10.1007/978-981-13-9193-4.

52. Stephan, M. *et al.* Radar Image Reconstruction from Raw ADC Data using Parametric Variational Autoencoder with Domain Adaptation. in *2020 25th International Conference on Pattern Recognition (ICPR)* (IEEE, 2021).

53. Cortes, C. & Vapnik, V. Support-vector networks. *Mach. Learn.* **20**, 273–297 (1995).

54. Chih-Wei Hsu & Chih-Jen Lin. A comparison of methods for multiclass support vector machines. *IEEE Trans. Neural Netw.* **13**, 415–425 (2002).

55. Labrie-Boulay, I. *et al.* Machine learning-based spin structure detection. (2023) doi:10.48550/ARXIV.2303.16905.

56. Schnitzspan, L., Kläui, M. & Jakob, G. Nanosecond True-Random-Number Generation with Superparamagnetic Tunnel Junctions: Identification of Joule Heating and Spin-Transfer-Torque Effects. *Phys. Rev. Appl.* **20**, 024002 (2023).

# Supplementary Information

## S1. Energy Efficiency

To estimate the energy consumption of a single device we examine the energy us of a scaled-down realistic version with a side length of 400 nm and a skyrmion size of 40 nm based on existing skyrmions stable at room temperature[1]. At these dimensions, the ratio between the skyrmion and the device closely resembles that of the experimentally used 36 $\mu$m sized device. The device's resistance is calculated to be approximately $R = 1$ k$\Omega$ between the two connected contacts[2].

With a current density required to move the skyrmion of $J = 2.9 \times 10^7$ Am$^{-2}$ at half the triangle's width, we find a flow of up to $I = 64$ nA through the device, including the 5 nm tantalum capping layer. Notably, as the capping layer serves as an anti-oxidation layer and does not impact the skyrmion dynamics, nearly half of the utilized current could be by either connecting the device from the bottom or altering the capping layer material. Given these parameters, the worst-case power consumption of one module would be $P = RI^2 = 4$ pW, assuming continuous application of maximum voltage, while typically only a fraction of this is used in a typical signal.

Changing the skyrmion size also affects its dynamics and therefore reducing the time needed per input. On the one hand, the diffusion constant increases with smaller skyrmions[3] and on the other hand, the required path to travel becomes shorter.

To obtain the energy for the operation we first consider conservatively the low diffusion constant in the case of significant pinning as shown in the main text. This assumption yields a time per gesture of $t = 1.1$ ms. Consequently, we calculate an energy consumption of $E = Pt = 4$ fJ per gesture. Which is comparable to conventionally used CMOS ring oscillators[4].

The free diffusion expected theoretically[5] and that we have shown experimentally can be generated by AC excitations[6] yields for these skyrmions a diffusion constant of $70 \times 10^{-9}$ m$^2$s$^{-1}$ leading to a reduced time per operation of 780 ns, resulting in ultra-low energy consumption of 3 aJ per gesture, significantly lower than the scaled down version of a previously proposed spin-torque nano-oscillators at 100 aJ[7].

While we show that our device compares favorably to other reservoir computing devices, for the overall power consumption one needs to also consider the CMOS periphery. We emphasize that our device operates in quasi-dc regime and therefore has reduced demands for CMOS periphery as compared to oscillator based AC magnetic devices[7]. Hence, we expect that the CMOS overhead will reduce compared to the previously analyzed peripheral circuitry. For details of this power consumption, we refer to the pertinent literature[7,8].

In our architecture, the final layer incorporates a linear support vector machine (SVM). Recent progress has enabled the integration of SVMs into Field-Programmable Gate Arrays (FPGAs), facilitating an even more power-efficient implementation[9]. We note that a comparison with purely software-based solutions is not easily possible due to the complexity of now-a-days used CPUs or Cloud solutions and such an analysis comprises research that goes clearly beyond the scope of this work.

## Supplementary References

1. Everschor-Sitte, K., Masell, J., Reeve, R. M. & Kläui, M. Perspective: Magnetic skyrmions—Overview of recent progress in an active research field. *J. Appl. Phys.* **124**, 240901 (2018).

2. Raab, K. *et al.* Brownian reservoir computing realized using geometrically confined skyrmion dynamics. *Nat. Commun.* **13**, 6982 (2022).

3. Zázvorka, J. *et al.* Thermal skyrmion diffusion used in a reshuffler device. *Nat. Nanotechnol.* **14**, 658–661 (2019).

4. Saheb, Z., El-Masry, E. & Bousquet, J.-F. Ultra-low voltage and low power ring oscillator for wireless sensor network using CMOS varactor. in *2016 IEEE Canadian Conference on Electrical and Computer Engineering (CCECE)* 1–5 (2016). doi:10.1109/CCECE.2016.7726620.

5. Miltat, J., Rohart, S. & Thiaville, A. Brownian motion of magnetic domain walls and skyrmions, and their diffusion constants. *Phys. Rev. B* **97**, 214426 (2018).

6. Gruber, R. *et al.* 300-Times-Increased Diffusive Skyrmion Dynamics and Effective Pinning Reduction by Periodic Field Excitation. *Adv. Mater.* **35**, 2208922 (2023).

7. Romera, M. *et al.* Vowel recognition with four coupled spin-torque nano-oscillators. *Nature* **563**, 230–234 (2018).

8. Mizrahi, A. *et al.* Neural-like computing with populations of superparamagnetic basis functions. *Nat. Commun.* **9**, 1533 (2018).

9. Afifi, S., GholamHosseini, H. & Sinha, R. FPGA Implementations of SVM Classifiers: A Review. *SN Comput. Sci.* **1**, 133 (2020).